\documentclass[twocolumn]{article}

\usepackage{titling}

\usepackage{hyperref}
\hypersetup{
    colorlinks=true,
    linkcolor=blue,
    filecolor=red,      
    urlcolor=blue,
    citecolor=blue,
  }

\pretitle{
    \begin{center}
    \fbox{\begin{minipage}{0.8\textwidth}
        \centering \footnotesize \copyright\ 2026 IEEE Software (to Appear)\\  \url{https://doi.org/10.1109/MC.2026.3666634}. \\ All rights reserved. \\ Posted with permission
    \end{minipage}}
    \end{center}
    \vspace{2cm} 
    \begin{center}\LARGE
}
\posttitle{\end{center}}

\title{Making Software Metrics Useful}
\author{Ewan Tempero, Paul Ralph}
\begin{document}

\maketitle
\begin{abstract}
Most engineers use measurements to make decisions. However, measurements are rarely used for decisions about constructing software products.  While many approaches to measuring attributes of software (``metrics'') have been developed, they are rarely used to answer useful questions such as ``Do I need to refactor this class?'' or ``Are these integration tests sufficient?''  Practitioners therefore question the value of software metrics.  We argue that this situation arose because software metrics were developed without understanding metrology (the science of measurement) and suggest directions software metrics research should take. 
\end{abstract}

\section*{Column}

Software engineers do not seem to use measurements to make decisions about constructing software systems as often as other kinds of engineers use measurements to make decisions about constructing mechanical, electrical (etc.) systems---measurements that form the foundation of
successful projects~\cite{beitz1996engineering}.

While many means of measuring software (``software metrics'') have been created
and are used in research, developers rarely use them to make
decisions such as whether to refactor a class or which classes to create. This suggests that decision makers do not find existing
software metrics useful. In this column,
we discuss the problems with existing metrics, and suggest directions
software metrics research should take.

To clarify, ``\textit{Measurement} is the assignment of [values] to objects or events according to rules; \dots a \textit{measure} is the [value] assigned; \dots [and] a \textit{metric} is a method, algorithm, or procedure for assigning [values] to a phenomenon'' \cite{ralph2024teaching}. We do not measure \emph{entities} (e.g. humans, or functions); we measure \emph{attributes} of entities (e.g. \emph{height}, \emph{number of decision points}). Some attributes seem straightforward to measure (e.g. \emph{height}, \emph{decision points}) while others do not (e.g. \emph{health}, \emph{complexity}).
Any discussion of software metrics must explicate both the \textit{attributes} and the \textit{entities} being measured (e.g. complexity of a method vs. size of a system).

This column focuses on \emph{code quality
  attributes} (QAs): traits related to the goodness of computer instructions such as
complexity, modifiability, understandability, or testability. Discussions
about how to improve software quality usually refer to QAs such as these,
directly or indirectly (e.g. \cite{Parnas:CACM:1972,
  GammaHelmJohnsonVlissides:BOOK:1994, Fowler:BOOK:1999}). 
  
However, we must distinguish between the attribute we want to
measure, and the attribute a given metric actually measures. For example, Cyclomatic Complexity (CC) has been presented as a metric for ``complexity'' \cite{McCabe:TSE:1976} but what CC actually measures is number of linearly independent paths, which is at most one facet of software complexity. If a developer is trying to decide whether to refactor a \textit{function} (the entity) that seems too \textit{complex} (the attribute), CC is of limited help because the number of linearly independent paths only tells one part of a multifaceted story. 

Psychologists encountered similar problems in predicting human behavior: individual metrics (e.g. personality survey questions) were insufficient. To make accurate predictions, they need holistic estimates of second-order ``constructs'' (AKA latent variables), each of which is estimated by aggregating multiple metrics (called \emph{indicators})
that together support decision making~\cite{Cronbach:PB:1955}. This ``reflective'' approach to measurement is now common across many social, applied, and natural sciences. 

In contrast, SE researchers often seek ``the best'' single metric, proposing new metrics and arguing they are better than previous ones. These new metrics are often slight variations on previous ones. The result is a set of competing metrics, none of which capture the breadth of the (often complicated, multidimensional) attribute we care about. These oversimplified metrics distort reality, undermining decision-making.

Psychologists do not seek the best individual metric. Psychologists seek the best \textit{measurement model}: a constellation of metrics and a way of aggregating them to estimate a construct. Multifaceted constructs demand different metrics, each of which captures a different facet. 

A measurement model for complexity, then, might include CC as well as indicators of level of
nesting, the complexity of conditional expressions, and other facets of complexity. Rather than create a new, overcomplicated and hard-to-understand super-complexity
metric that counts linearly independent paths, nesting levels, etc., we should use multiple metrics (indicators), each providing different information about the construct, and aggregate them using well-understood statistical approaches (e.g. factor analysis, not average). 

To choose good indicators---indeed to get anywhere with measurement---we need to abandon the outdated idea that we can observe and count everything that matters. Rather, sociotechnical phenomena (such as software engineering) are largely determined by latent structures (e.g. code quality, projects, technical debt). Once we recognize that these latent structures are real and affect our world, we can go about creating appropriate measurement models that operationalize latent variables as the shared variance of \textit{multiple} indicators (metrics), all of which correspond to the various facets of the target latent variable. 

To determine whether a metrics is valid, we need to clarify \textit{why} we measure things: to \textit{compare} them.
We can avoid measurement by comparing directly. For example, to cut a wooden board for a gap in a floor, we can simply lay the board on the floor, line it up, and mark where to cut without ever quantifying the board's length. Unfortunately, software usually cannot be directly compared in this way so we have to measure. 

Comparing
measurements should always give the same result as direct comparison; that
is, the relationship between the measurements of an attribute for two entities
is the same as the directly observed relationship between the entities with
respect to the attribute.

One criterion for a ``good'' metric is therefore that
the measurements given by the metric are related in the same way as the
attributes of the entities being measured---what Fenton and Pfleeger called the \emph{representation condition}
\cite{FentonPfleeger:BOOK:1997}.

For example, consider measuring the quality of a flower garden. We observe
that gardens with rocks in them tend to be harder to manage, so we conclude we
can measure the quality of gardens by counting the number of rocks. In order
for this metric to satisfy the representation condition, given two gardens,
the one with the fewest rocks should be considered the better garden.  

Clearly
there are issues with this. Rocks come in all shapes and sizes. Rocks may
overlap with each other, or there could be large gaps between. Rocks with the
right chemical composition may actually be better for the garden. Note that
using ``rock density'' instead does not change the usefulness of the metric. So this metric does not meet the representation condition. 
While this may seem a patently absurd example, it shares many characteristics
with proposed software metrics!

It is easy to find pairs of functions where the one with the higher CC seems ``less complex'' on direct comparison---so CC fails the representation condition \emph{for the attribute complexity}. It is perfectly fine for measuring the attribute of number of linearly independent paths. However CC does \emph{correlate} with complexity, and therefore is a good candidate for inclusion in the set of reflective indicators we would aggregate to estimate complexity.

So we need multiple metrics that measure things that change in response to changes in the quality attribute we care about. The question now is how to we determine what
is ``relevant''?  Essentially, the metrics need to correlate with the quality
attribute, and with each other. This is known as convergent validity. This is
not a simple process. We suggest how it can be done in previous work
\cite{RalphTempero:EASE:2018}.

In conclusion, we, as a research community, must abandon the idea of
a single best metric for a quality attribute. Any proposal for a new metric must:
\begin{itemize}
\item Clearly define the metric's target construct (e.g. complexity), the entities to which the metric applies (e.g. functions) and the specific attribute that it \emph{actually} measures (e.g. number of linearly independent paths) without conflating the attribute with the construct.
\item Demonstrate that the proposed metric meets the
  representation condition (often being explicit about the attribute will be sufficient, such as with linearly independent paths).
\item Provide a convincing argument (i.e. by discussing face and content validity) that the proposed metric \textit{should} correlate with the construct (e.g. the more paths through the code, the harder it will be to understand or test).
\item Provide empirical evidence that the proposed metric correlates more strongly with other
  metrics for the same quality attribute than with metrics for other, related quality attributes (e.g. a new complexity metric should correlate better with other complexity metrics than with size metrics)---AKA convergent and discriminant validity.
\end{itemize}

This will not be easy. Looking for useful indicators of a construct changes our understanding of the construct, which changes what it means to be a useful indicator. We should therefore expect code quality metrics and our  understanding of code quality itself to co-evolve. However, embracing comprehensive measurement models and better construct validation, as explained above, will help researchers create more sound and accurate measurement models, which will be more useful for informing software professional's decisions.

\bibliographystyle{plain}
\bibliography{bib}

@article{beitz1996engineering,
  title={Engineering design: a systematic approach},
  author={Beitz, W and Pahl, G and Grote, K},
  journal={{MRS} Bulletin},
  volume={71},
  pages={30},
  year={1996},
  publisher={Cambridge University Press Cambridge, UK}
}

@article{Cronbach:PB:1955,
author = {Cronbach, Lee J and Meehl, Paul E},
title = {{Construct validity in psychological tests}},
journal = {Psychological Bulletin},
year = {1955},
volume = {52},
number = {4},
pages = {281--302},
publisher = {American Psychological Association},
doi = {10.1037/h0040957},
language = {English},
uri = {\url{papers3://publication/doi/10.1037/h0040957}}
}

@Book{FentonPfleeger:BOOK:1997,
  author =       {Fenton, Norman E. and Pfleeger, Shari Lawrence},
  title =        {Software Metrics: A Rigorous \& Practical Approach},
  publisher =    {PWS Publishing Company},
  year =         1997,
  edition =      {Second}
}

@book{Fowler:BOOK:1999,
 author = {Martin Fowler},
 title = {Refactoring: improving the design of existing code},
 year = {1999},
 isbn = {0-201-48567-2},
 publisher = {Addison-Wesley},
 address = {Boston, MA, USA},
 }

@book{GammaHelmJohnsonVlissides:BOOK:1994,
 author = {Erich Gamma and Richard Helm and Ralph Johnson and John Vlissides},
 title = {Design Patterns},
 year = {1994},
 isbn = {0-201-63361-2},
 publisher = {Addison Wesley Publishing Company},
 address = {One Jacob Way, Reading, Massachusetts 01867},
}

@article{McCabe:TSE:1976,
        author = {Mccabe, T. J. },
        booktitle = {Software Engineering, IEEE Transactions on},
        journal = {Software Engineering, IEEE Transactions on},
        number = {4},
        pages = {308--320},
        title = {A Complexity Measure},
        url = {http://ieeexplore.ieee.org/xpls/abs_all.jsp?arnumber=1702388},
        volume = {SE-2},
        year = {1976}
}

@article{Parnas:CACM:1972,
 author = {D. L. Parnas},
 title = {On the criteria to be used in decomposing systems into modules},
 journal = {Commun. ACM},
 volume = {15},
 number = {12},
 year = {1972},
 issn = {0001-0782},
 pages = {1053--1058},
 doi = {http://doi.acm.org/10.1145/361598.361623},
 publisher = {ACM},
 address = {New York, NY, USA},
}

@incollection{ralph2024teaching,
  title={Teaching Software Metrology: The Science of Measurement for Software Engineering},
  author={Ralph, Paul and Kuutila, Miikka and Arif, Hera and Ayoola, Bimpe},
  booktitle={Handbook on Teaching Empirical Software Engineering},
  pages={101--154},
  year={2024},
  publisher={Springer}
}

@inproceedings{RalphTempero:EASE:2018,
  author = {Ralph, Paul and Tempero, Ewan},
  title = {Construct Validity in Software Engineering Research and Software Metrics},
  booktitle = {22nd International Conference on Evaluation and Assessment in Software Engineering},
  month = jun,
  year = {2018},
  url = {https://doi.org/10.1145/3210459.3210461},
}

\end{document}